\documentclass[%
 reprint,
 floatfix,
 twocolumn,
 amsmath,amssymb,
 aps,
]{revtex4-1}

\usepackage{amsmath}
\usepackage{graphicx}
\usepackage{dcolumn}
\usepackage{bm}
\usepackage{hyperref}

\usepackage{braket}
\usepackage{xcolor}
\usepackage{ulem}

\begin{document}

\preprint{APS/123-QED}

\title{Orbital Hall effect in crystals: inter-atomic versus intra-atomic contributions}

\author{Armando Pezo}
 \affiliation{Aix-Marseille Universit\'e, CNRS, CINaM, Marseille, France.}
\author{Diego Garc\'ia Ovalle}
 \affiliation{Aix-Marseille Universit\'e, CNRS, CINaM, Marseille, France.}
\author{Aur\'elien Manchon}%
\email{aurelien.manchon@univ-amu.fr}
\affiliation{Aix-Marseille Universit\'e, CNRS, CINaM, Marseille, France.}


\date{\today}

\begin{abstract}
The orbital Hall effect (OHE) designates the generation of a charge-neutral flow of orbital angular momentum transverse to an initial charge current. Recent theoretical investigations suggest that transition metals display sizable OHE, encouraging experimental search along this direction. Nonetheless, most of these theories assume that the orbital moment originates from the region immediately surrounding the atom core, adopting the so-called {\it atomic center approximation}. In periodic crystals though, the contribution of the interstitial regions is crucial and can lead to a severe misestimation of the OHE. By applying the "modern theory" of orbital magnetization to the OHE, we assess the relative importance of intra-atomic and inter-atomic contributions in selected materials from first principles. We find that whereas the OHE is mostly of intra-atomic origin for wide band-gap semiconductors (e.g., MoS$_2$), the inter-atomic contribution becomes crucial in narrow band-gap semiconductors (SnTe, PbTe) and transition metals (Pt, V etc.). These predictions invalidate the atomic center approximation adopted in some of the previous works and open perspectives for the realization of efficient sources of orbital currents.

\end{abstract}

                              
\maketitle

\noindent
{\it Introduction} - The need for energy-efficient microelectronic solutions has accelerated the efforts to identify degrees of freedom that could complement or replace the electron's charge to carry and store information. Whereas spintronics, which uses the electron's spin angular momentum (SAM) to transmit and manipulate data, is probably the most mature alternative technology to date \cite{Vedmedenko2020}, other directions have emerged in the past two decades seeking to exploit magnons in magnetic insulators \cite{Chumak2015} or the valley degree of freedom in certain low symmetry semiconductors \cite{Vitale2018}. Under these various paradigms, the charge of the electron is replaced by a quantum degree of freedom (SAM or valley) that survives in the semiclassical limit and may encode the information over two distinct values (spin up/down, valley K/K' etc.). In recent research, the control of the SAM is achieved via spin-orbit coupling, a property that scales with the mass of the elements nucleus. Therefore, most progress is currently achieved using heavy materials such as Pt, W, Bi etc. which are scarce and expensive \cite{Manchon2019}. Alternatively, the emergent field of valleytronics exploits valley-polarized currents induced by light in optically active materials. Whether the valley degree of freedom could be injected in adjacent materials, transported over long distances and stored remains an active area of research. In this context, the orbital angular momentum (OAM) has started to emerge as a promising degree of freedom that could be generated efficiently and transported over long distances \cite{Bernevig2005b,Go2018,Hayashi2022}.

Equilibrium OAM has been investigated thoroughly over the past decade and was shown to substantially contribute to the overall magnetization in certain classes of time-reversal broken materials, associated with the ground state's Berry curvature \cite{Thonhauser2005,Shi2007}. In fact, equilibrium OAM necessitates time-reversal symmetry breaking combined with either non-collinear magnetic texture or spin-orbit coupling (hence, heavy metal elements) \cite{Hanke2016,Hanke2017}. Although equilibrium OAM vanishes when time-reversal symmetry is preserved, in the presence of an external electric field nonequilibrium OAM can be generated even in the absence of spin-orbit coupling, as long as inversion symmetry is broken \cite{Yoda2018,Jo2018}. This effect is tagged "orbital Rashba-Edelstein effect" (ORE), in analogy to the celebrated spin Rashba-Edelstein effect (SRE) that enables the electrical generation of SAM \cite{Edelstein1990,Manchon2015}. Similar to SRE, ORE features an electron density that carries OAM and is limited to systems lacking inversion symmetry such as interfaces and noncentrosymmetric crystals. The OAM is not {\it transported} through the crystal though, it is rather created {\it locally}, which undermines its application to nonlocal orbitronics devices.

In contrast, in centrosymmetric crystals charge-neutral flows of OAM can be induced by electrical field via the orbital Hall effect \cite{Bernevig2005b} (OHE), the orbital analog to the spin Hall effect \cite{Sinova2015} (SHE). Theories predict that OHE is usually much larger than SHE \cite{Kontani2009}, a feature that is particularly striking in light metals \cite{Jo2018,Salemi2022}. These theoretical works have inspired recent experimental observations such as the orbital torque \cite{Go2020,Ding2020,Lee2021b} and the orbital magnetoresistance \cite{Ding2022}. Nonetheless, these calculations assume that the OAM originates from the region immediately surrounding the atom core, adopting the so-called {\it atomic center approximation} (ACA). In periodic crystals though, the OAM does not only arise from the atom core but also from the interstitial, inter-atomic region as accounted for by the "modern theory" of orbital magnetization \cite{Thonhauser2005,Shi2007}. In fact, it was recently shown that OHE solely originates from {\it inter-atomic} contribution in gapped graphene \cite{Bhowal2021}, whereas both intra-atomic (atom core) and inter-atomic (interstitial) terms are of comparable magnitude in MoS$_2$ bilayers \cite{Cysne2022}.

In this Letter, we assess the relative magnitude of the intra- and inter-atomic contributions to the OHE in selected materials of prime importance to experiments. Whereas the ACA tends to be mostly valid in large band-gap semiconductors (e.g., MoS$_2$ monolayer), it fails in both narrow band-gap semiconductors (SnTe, PbTe) and transition metals (V and Pt). This finding suggests that previous estimates \cite{Kontani2009,Jo2018,Salemi2022} need to be revisited by accounting for the total contribution to the OAM \cite{Bhowal2021,Cysne2022}. Due to their different character, local (atom core) and nonlocal (interstitial), the intra-atomic and inter-atomic contributions are expected to behave differently in the presence of disorder and to play a distinct role in orbital torque and pumping \cite{Go2020}.

{\it Orbital Hall conductivity} - In the linear response theory the OAM current is time-reversal symmetric, akin to the spin current, and is governed by the intrinsic Fermi sea term of the Kubo formula \cite{Bonbien2020}. In other words, in the limit of weak momentum scattering, the OAM conductivity reads \cite{Kontani2009}
\begin{equation}
        \sigma_{ij}^{\gamma}=-e\int_{BZ}\frac{d^3{\bf k}}{(2\pi)^3}\sum_{n}f_n(\mathbf{k})\Omega_{n,ij}^{o,\gamma}(\mathbf{k}),\label{i1}
\end{equation}
where the orbital Berry curvature is
\begin{equation}
        \Omega_{n,ij}^{o,\gamma}(\mathbf{k})=2\hbar \operatorname{Im}\sum_{m\not=n} \frac{\braket{u^n_{\bf k}|\mathcal{J}_{o,i}^\gamma|u^m_{\bf k}}\braket{u^m_{\bf k}|\hat{v}_j|u^n_{\bf k}}}{(\varepsilon^n_{\bf k}-\varepsilon^m_{\bf k})^2}.\label{i2}
\end{equation}
Here, $|u^n_{\bf k}\rangle$ is the periodic part of the Bloch state associated with the energy $\varepsilon^n_{\bf k}$. In addition, $f_n(\mathbf{k})$ is the equilibrium Fermi distribution function and $\mathbf{\hat{v}}=\hbar^{-1}\partial_{\mathbf{k}} \mathcal{H}_{\mathbf{k}}$ is the velocity operator, $\mathcal{H}_{\mathbf{k}}$ being the Hamiltonian in momentum space. The orbital current operator is defined as $\mathcal{J}_{o,i}^\gamma=\{\hat{v}_i,\hat{L}_\gamma\}/2$, where $\hat{\bf v}$ is the velocity operator and $\hat{\bf L}=\hat{\bf r}\times\hat{\bf p}$ is the OAM operator in the unit of the Planck constant $\hbar$. The indices $i$ and $\gamma$ denote the flow and orbital polarization directions, respectively, $\hat{\bf p}$ is the momentum of the carrier wave packet and $\hat{\bf r}$ represents its absolute position in the laboratory frame. Since the position operator $\hat{\bf r}$ is not well defined in the usual Bloch state representation, Ref. \cite{Thonhauser2005} showed that by adopting the Wannier representation, the orbital angular momentum operator can be parsed into two terms, $\hat{\bf L}=\bar{\bf r}_i\times\hat{\bf p}+(\hat{\bf r}-\bar{\bf r}_i)\times\hat{\bf p}$, with $\bar{\bf r}_i$ being the position of the Wannier center. The first term is associated to the local current circulation, in the vicinity of the atom core, and the second term is associated with the nonlocal circulation in the interstitial space between the atoms. Within the ACA, only the first term is considered while the second one is neglected \cite{Kontani2009,Jo2018,Salemi2022}; as a result, the intra-atomic orbital current operator $\mathcal{J}_{o,i}^\gamma|_{\rm intra}$ is straightforwardly defined in the basis of spherical harmonics of each atom in the unit cell.

The modern theory of orbital magnetization \cite{Thonhauser2005,Shi2007} does not perform the aforementioned separation, and rather expresses the total OAM operator in terms of the crystal Bloch states, properly accounting for the corrections due to Berry connection. The modern theory has been investigated by first principles calculations, and it was found that whereas the ACA is qualitatively valid for bulk insulating transition-metal oxides and magnetic transition metals (Co, Fe, Ni), it substantially fails at interfaces \cite{Nikolaev2014,Hanke2016}. As mentioned above, extending the theory of OHE beyond the ACA \cite{Bhowal2021} is indispensable to the design of experiments and devices based on orbital transport. In Eq. \eqref{i2}, the interband matrix element reads \cite{Bhowal2021}
\begin{eqnarray}
\braket{u^n_{\bf k}|\mathcal{J}_{o,i}^\gamma|u^m_{\bf k}}=\frac{1}{2}\sum_p&\left(\langle u^n_{\bf k}|\hat{v}_i|u^p_{\bf k}\rangle\langle u^p_{\bf k}|\hat{L}_\gamma|u^m_{\bf k}\rangle\right.\nonumber\\
&\left.+\langle u^n_{\bf k}|\hat{L}_\gamma|u^p_{\bf k}\rangle\langle u^p_{\bf k}|\hat{v}_i|u^m_{\bf k}\rangle\right).\label{eq3}
\end{eqnarray}
Using the (symmetrized) definition of the OAM, $\hat{\bf L}=(\mathbf{\hat{r}}\times \mathbf{\hat{p}}-\mathbf{\hat{p}}\times \mathbf{\hat{r}})/4$, one obtains \cite{SuppMat,Bhowal2021}
\begin{eqnarray}
\langle u^n_{\bf k}|\hat{\bf L}|u^p_{\bf k}\rangle=&\frac{e}{2g_L\mu_B}{\rm Im}\langle \partial_{\bf k}u^n_{\bf k}|\times\mathcal{H}_{\bf k}|\partial_{\bf k}u^p_{\bf k}\rangle\nonumber\\
&-\frac{e}{4g_L\mu_B}(\varepsilon^n_{\bf k}+\varepsilon^p_{\bf k}){\rm Im}\langle \partial_{\bf k}u^n_{\bf k}|\times|\partial_{\bf k}u^p_{\bf k}\rangle.
\end{eqnarray}
Here, $\mu_B=e\hbar/2m_e$ is Bohr's magneton and $g_L$ is the orbital g-factor. Following the estimates of Ref. \cite{MacDonald1982} for transition metals, we adopt $g_L$=1. By considering $|\partial_{\bf k}u^n_{\bf k}\rangle=\hbar\sum_{q\neq n}\frac{\langle u^q_{\bf k}|\hat{\bf v}|u^n_{\bf k}\rangle}{\varepsilon^q_{\bf k}-\varepsilon^n_{\bf k}}|u^q_{\bf k}\rangle$, one finally deduces that
\begin{align}
     \langle u^n_{\bf k}|\hat{\bf L}|u^p_{\bf k}\rangle&=\frac{e\hbar^2}{4\mu_B}\operatorname{Im}\sum_{q\not=n,p}\left(\frac{1}{\varepsilon^q_{\bf k}-\varepsilon^n_{\bf k}}+\frac{1}{\varepsilon^q_{\bf k}-\varepsilon^p_{\bf k}}\right)\nonumber\\
     &\braket{u^n_{\bf k}|\mathbf{\hat{v}}|u^q_{\bf k}}\times\braket{u^q_{\bf k}|\mathbf{\hat{v}}|u^p_{\bf k}}.\label{i3}
\end{align}
Inserting Eq. \eqref{i3} into Eq. \eqref{eq3}, one infers the expression of the total OHE, including both intra- and inter-atomic contributions. It is important to emphasize a key difference between the intra-atomic and total OHE expressions. Whereas both effects are inversely proportional to $(\varepsilon^n_{\bf k}-\varepsilon^m_{\bf k})^2$ [Eq. \eqref{i2}], suggesting hot spots close to avoided band crossing in the Brillouin zone, the total orbital current is additionally influenced by a factor proportional to the relative energy difference between bands, $\varepsilon^n_{\bf k}-\varepsilon^m_{\bf k}$ [Eq. \eqref{i3}]. Therefore, given the relation between intra-atomic and total responses, we expect the inter-atomic OHE to be more sensitive to the band ordering than the intra-atomic OHE, which could lead to cancellation when numerous bands are involved (e.g., in a transition metal) or when disorder is present. We now evaluate these contributions in selected examples.

{\it Minimal model for OHE} - To evaluate the impact of the aforementioned OHE contributions, we first select a minimal model with a restricted set of spinless atomic orbitals (typically, $p_x$ and $p_y$). An example of such a toy model was introduced by Fu \cite{Fu2011} as a paradigm for topological crystalline insulators. From our standpoint, this model presents the advantage that its topological properties are due to the OAM character of the Bloch states rather than to their SAM character. It consists of a square bipartite lattice whose Hamiltonian reads
\begin{eqnarray}
   {\cal H} &=& \sum_n {\cal H}^A_n+{\cal H}^B_n+{\cal H}_n^{AB},\label{5} 
\end{eqnarray}
where the Hamiltonian for each sublattice $ {\cal H}^{a}_n$, $a=A,B$, and the coupling Hamiltonian $ {\cal H}^{AB}_n$ are given by
\begin{eqnarray}
 {\cal H}^{a}_n &=&\sum_{i,j}t^{a}(\mathbf{r}_i-\mathbf{r}_j)\sum_{\alpha,\beta}c^{\dagger}_{a,\alpha}(\mathbf{r}_i,n)e^{i,j}_{\alpha}e^{i,j}_{\beta}c_{a,\beta}(\mathbf{r}_j,n),\label{6}\\
    {\cal H}^{AB}_n&=&\sum_{i,j}t'(\mathbf{r}_i-\mathbf{r}_j)\left[ \sum_{\alpha} c^{\dagger}_{A,\alpha}(\mathbf{r}_i,n)c_{B,\alpha}(\mathbf{r}_j,n)+h.c. \right]\nonumber\\
&&+t_z\sum_i \sum_{\alpha}[c^{\dagger}_{A,\alpha}(\mathbf{r}_i,n)c_{B,\alpha}(\mathbf{r}_i,n+1)+h.c.].\label{7}
\end{eqnarray}
Each site is identified by the bilayer unit cell $n$, the planar coordinate in each layer $\mathbf{r}=(x,y)$, the sublattice label $a=A,B$ and the orbital index $\alpha,\;\beta$. The unit vectors ${\bf e}^{i,j}=(\mathbf{r}_i-\mathbf{r}_j)/|\mathbf{r}_i-\mathbf{r}_j|$ account for the directionality of the hopping integrals. We recall that this setup can represent a $p_{x,y}$ system just like a $d_{xz,yz}$ one since both transform in the same way under $C_4$. Without loss of generality, we limit our treatment to nearest neighbor ($t_1^a$, $t'_1$) and next-nearest neighbor hopping ($t_2^a$, $t'_2$).
\begin{figure}[ht!]
\includegraphics[width=\linewidth]{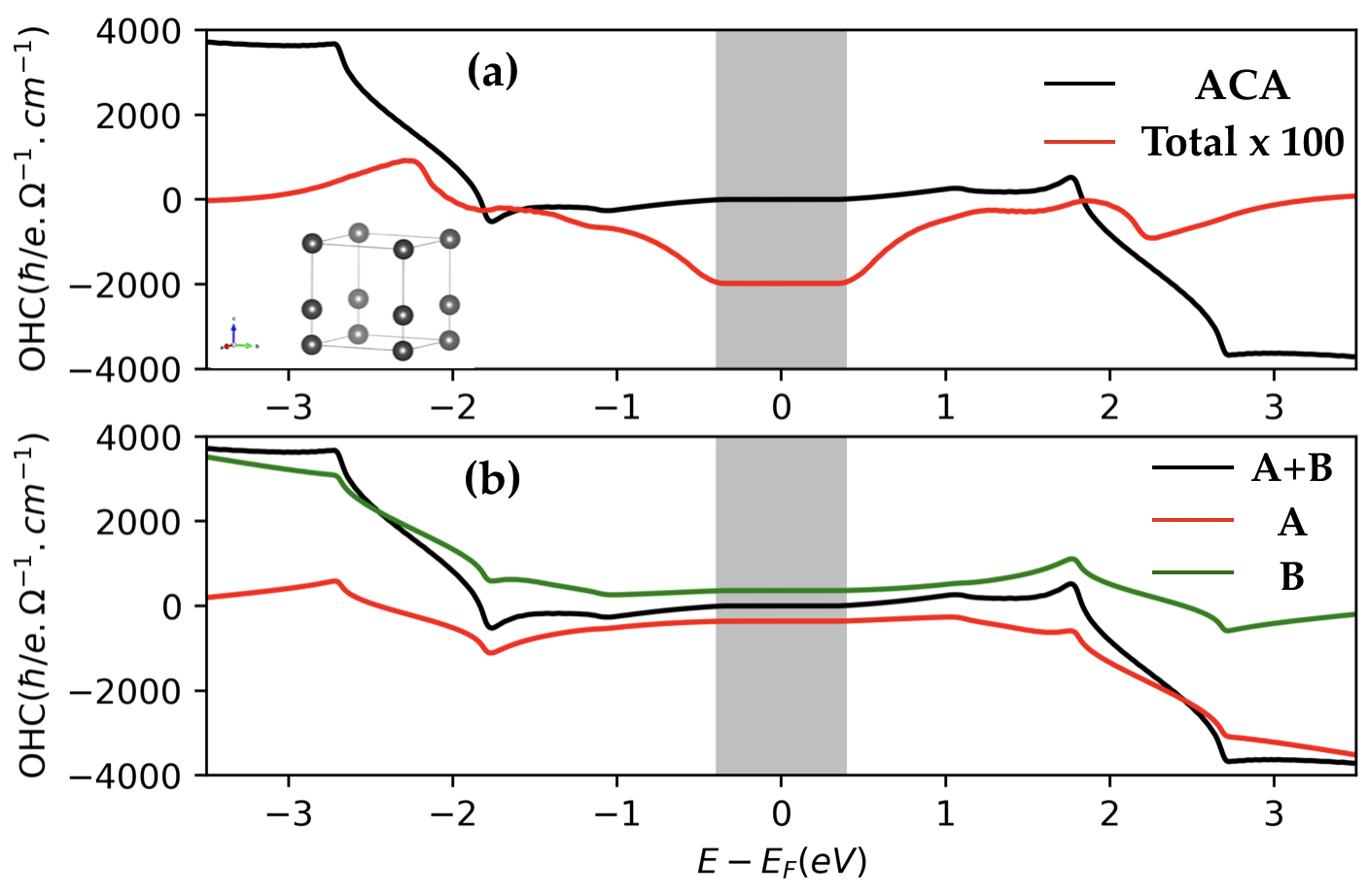}
\caption{(Color online) (a) Orbital Hall conductivity computed for the 4-band model. We compute both intra-atomic (ACA - black) and total contributions (red). (b) Intra-atomic OHE projected on sublattices A (red) and B (green). These projections are equal and opposite in the gap, leading to a vanishing overall intra-atomic OHE. The parameters are set to $t^a_1=-t^b_1=1,\;t^a_2=-t^b_2=0.5,\;t'_1=2.5,\;t'_2=0.5,\;t_z=2$. The inset displays the unit cell.}
\label{fig:ohcs_cti}
\end{figure}

In Fig. \ref{fig:ohcs_cti}, we show both intra-atomic and total OHE conductivities $\sigma^z_{xy}$ computed using Eq. \eqref{i1} as a function of the energy. We find that in the gap vicinity [grey shaded region in Fig. \ref{fig:ohcs_cti}(a)], the intra-atomic OHE vanishes (black line), whereas the total OHE is finite (red line) but not quantized. In other words, the OHE is of purely inter-atomic origin in spite of the non-vanishing atomic OAM character of the bands. Remarkably, when projected on each sublattice the intra-atomic OHE is quantized and staggered [red and green lines in Fig. \ref{fig:ohcs_cti}(b)]. In other words, Fu's model does not realize a quantum orbital Hall insulator, but rather a quantum 'staggered' orbital Hall insulator. The finite total OHE shown in Fig. \ref{fig:ohcs_cti}(a) is attributed to Berry curvature peaks appearing at certain points in the Brillouin zone, as shown in the Supplemental Material \cite{SuppMat}. Another important feature is that whereas inter-atomic OHE dominates in the gap, intra-atomic OHE increases away from the gap, which suggests that intra-atomic and inter-atomic OHE might be distinguishable in certain multiband systems.

{\it Realistic material simulations -} We now turn to the simulation of OHE in real materials. We start by considering well-known semiconductors that display strong orbital hybridization in their band structure. As paradigmatic narrow-gap semiconductors, we select the three-dimensional topological crystalline insulator SnTe \cite{Hsieh2012}, and its topologically trivial parent compound, PbTe, both of which possess large $p$ orbitals hybridization near the gap located at the L points in the Brillouin zone. As an example of a large-gap semiconductor, we chose MoS$_2$-2H monolayer. In fact, MoS$_2$ and transition metal dichalcogenide siblings possess two valleys at K and K' points in the Brillouin zone and support valley Hall effect \cite{Xiao2012b}, related to intra-atomic OHE \cite{Bhowal2020,Canonico2020,Cysne2021}. We perform density functional theory (DFT) simulations \cite{dft1964,dft1965} using the Perdew-Burke-Ernzerhof exchange-correlation functional \cite{gga,Perdew1996}. We achieved the geometry optimization with the plane-wave basis as implemented in the Vienna $\textit{Ab-initio}$ Simulation Package (VASP) \cite{Kresse1996b,Kresse1996a}. For SnTe and PbTe, we used a 400 eV cutoff for the plane-wave expansion along with a force criterion $<$ 5 $\mu$eV/\AA\;with a $15 \times 15 \times 15$ $\mathbf{k}$-points sampling of the Brillouin zone. The ionic potentials were described using the projector augmented-wave (PAW) method \cite{Kresse1999}. For MoS$_2$-2H monolayer, we used a 350 eV cutoff with 15$\times$15$\times$1 Monkhorst pack for the k-grid. Taking a 15{\AA} vacuum to avoid interaction with mirror images, the structure was relaxed such that the forces satisfied the criterion $<$ 10 $\mu$eV/\AA. 

In all three cases, the Hamiltonian matrix was obtained by Wannier interpolation as implemented in the Wannier90 package \cite{Pizzi2020}. For SnTe and PbTe, we have used the $s$ and $p$ orbitals which are responsible for the electronic properties of this material around the $L$ high symmetry point \cite{Ye2015b,Littlewood2010}, whereas for MoS$_2$-2H monolayer we have used a basis considering the transition metal $d$ orbitals along with the chalcogen $p$ orbitals. Finally, in each case we have symmetrized the real space Hamiltonian by imposing lattice symmetry constrains \cite{Zhi2022}. The bands structures are displayed in \cite{SuppMat}.

\begin{figure}[h]
\includegraphics[width=\linewidth]{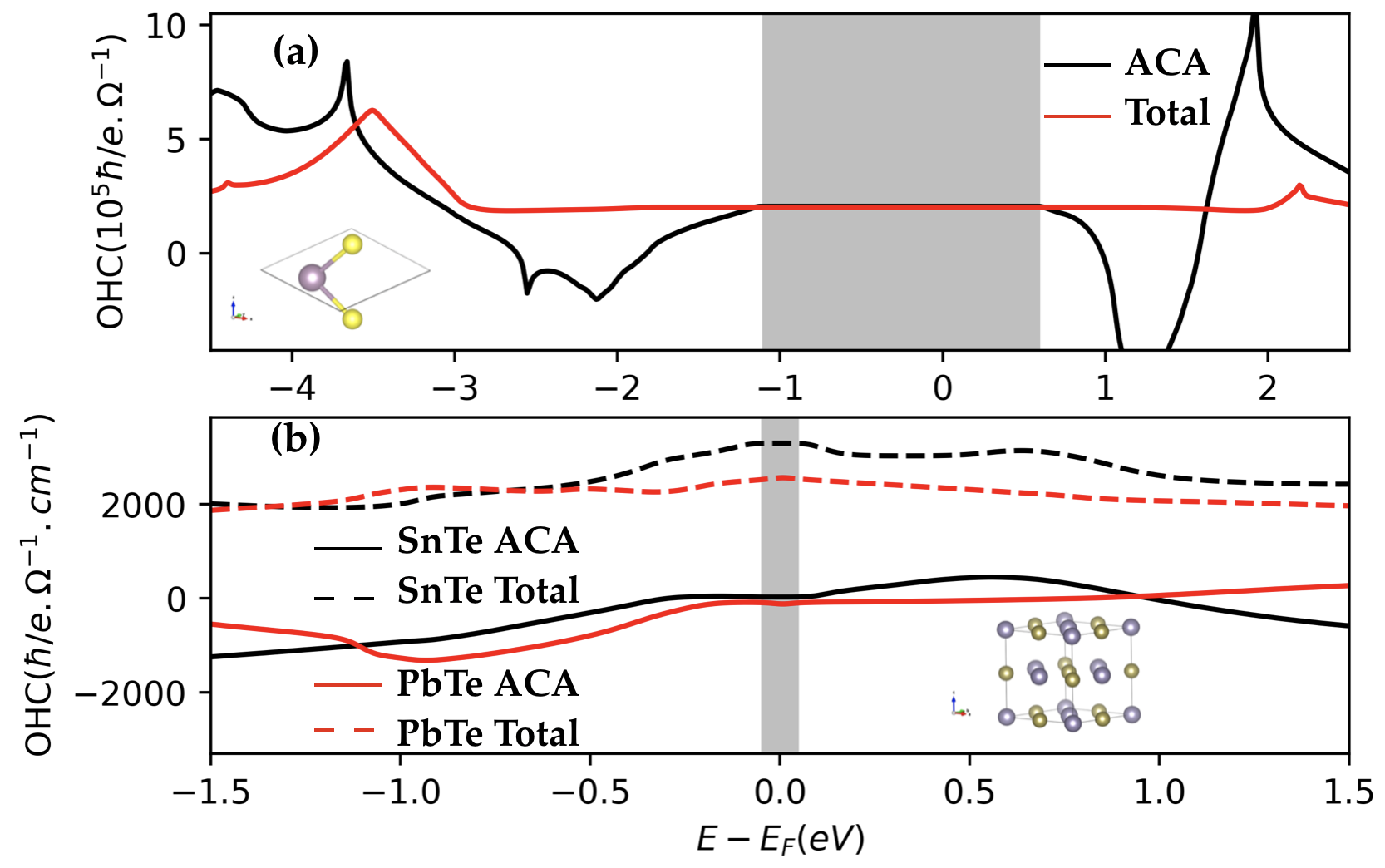}
\caption{(Color online) (a) Intra-atomic (ACA - black) and total (red) OHE for MoS$_2$-$2H$ monolayer. The inset displays the unit cell and the grey shaded region indicates the gap. ACA is valid in the vicinity of gap but fails away from it. (b) Intra-atomic (ACA - solid lines) and total (dashed lines) OHE for SnTe (black) and PbTe (red). The inset displays the unit cell and the grey shaded region indicates the gap. The failure of ACA suggests that the OHE possesses a dominant inter-atomic contribution.}
\label{fig:ohc_mos2}
\end{figure}

The intra-atomic and total OHE conductivities are reported in Fig. \ref{fig:ohc_mos2} for (a) MoS$_2$ monolayer, as well as for (b) SnTe and PbTe. In MoS$_2$, since each valley is associated with an OAM of opposite sign, one should expect that the valley Hall effect is accompanied by an OHE, as pointed out recently \cite{Bhowal2020,Canonico2020,Cysne2021}. In fact, in agreement with these studies, we obtain a finite value of the intra-atomic OHE in the gap for MoS$_2$ [black line in Fig. \ref{fig:ohc_mos2}(a)] coinciding with the value of the total OHE (red line), leaving a negligible inter-atomic contribution. This finite, but not quantized, value of the intra-atomic OHE in the gap reveals that the edge states lack robustness when disorder is included \cite{C9CP01590F}.

The narrow-gap semiconductors, SnTe and PbTe, offer a strikingly different picture as shown in Fig. \ref{fig:ohc_mos2}(b). SnTe has an inverted gap of $\sim$ 0.15 eV and the active region near the gap is mostly composed of $p$ orbitals. The same is true for PbTe although with a narrower gap ($\sim$ 0.08 eV) and no band inversion. We find that the total OHE (dashed lines) is much larger than the intra-atomic OHE, indicating that nonlocal contributions are crucial in both materials. In the case of SnTe, the intra-atomic OHE vanishes in the gap, whereas for PbTe, it is finite but small. We notice that the magnitude of the inter-atomic OHE scales inversely with the gap: it decreases from PbTe to SnTe and vanishes in MoS$_2$,  which reflects the progressive increase of the localization of the Wannier states in the gap region [see Eq. \eqref{i3}].

\begin{figure}[ht!]
\includegraphics[width=\linewidth]{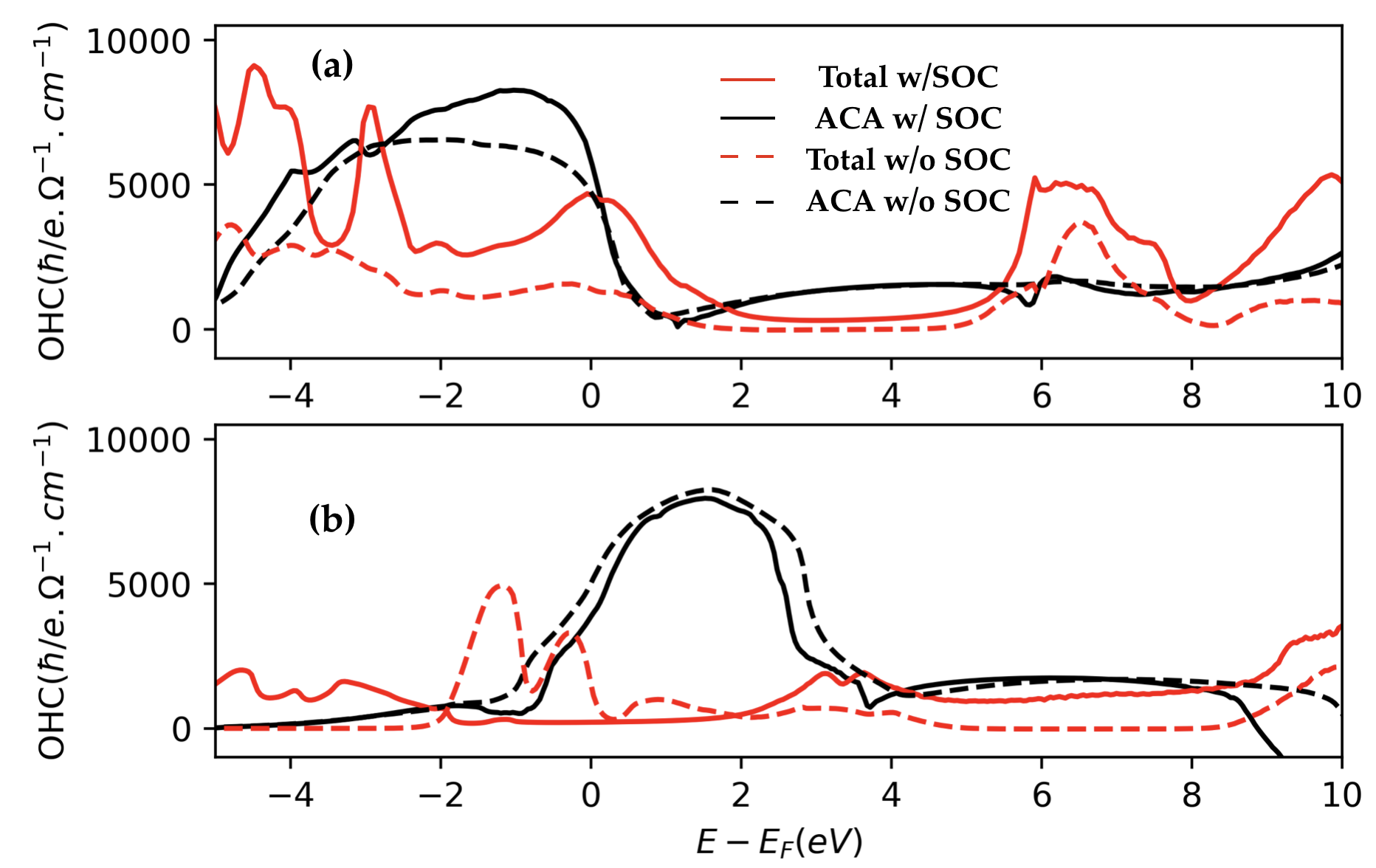}
\caption{(Color online) Intra-atomic (ACA - black) and total OHE (red) for bulk Pt (a) and bulk V (b), without (dashed line) and with (solid line) spin-orbit coupling. The total OHE is systematically smaller than the one estimated within ACA, pointing out the importance of inter-atomic contribution in transition metals. In addition, whereas spin-orbit coupling as a minor influence on the intra-atomic contribution (black), the total OHE (red) being more sensitive to band alignment, it is dramatically impacted by turning on spin-orbit coupling.}
\label{fig:ohc_v_pt}
\end{figure}

In the systems discussed so far, we have considered semiconductors whose band structure in the vicinity of the gap is reasonably modelled by a few bands only. We now move on to transition metals that display no gap and involve a large number of bands close to Fermi level. In Fig. \ref{fig:ohc_v_pt}, we show the intra-atomic and total OHE for two representative metallic materials with large [Pt, Fig. \ref{fig:ohc_v_pt}(a)] and weak spin-orbit coupling [V, Fig. \ref{fig:ohc_v_pt}(b)]. The values obtained for the intra-atomic OHE (black lines) are in good agreement with Ref. \cite{Jo2018}, but exceed by a large margin the total OHE (red lines). We remind that the total OHE is highly sensitive to the relative band alignment [see Eq. \eqref{i3}], which can lead to an overall cancellation of the total OHE in metals. The fact that inter-atomic and intra-atomic OHE contributions are of comparable magnitude in transition metals is surprising considering that the equilibrium orbital magnetization of bulk magnetic transition metals (Ni, Co and Fe) is mostly intra-atomic \cite{Hanke2016}. This is another illustration of the fact that nonequilibrium OHE is much more sensitive to band structure details than equilibrium orbital magnetization. Finally, let us comment on the impact of spin-orbit coupling on the OHE. As reported on Fig. \ref{fig:ohc_v_pt}, the intra-atomic OHE (black lines) is only weakly influenced by spin-orbit coupling whereas the total OHE (red lines) is again much more sensitive. As a result, great care should be taken when computing the OHE in bulk transition metals \cite{Kontani2009,Jo2018,Salemi2022}. The high sensitivity of the total OHE to band structure peculiarities might result in dramatic modifications upon interfacial tuning and electrical gating, opening routes to OHE engineering.

{\it Discussion} - The present study shows that the conventional ACA used to compute the equilibrium orbital magnetization \cite{Hanke2016} in transition metals is not appropriate when considering the OHE and should be replaced by the modern theory. In semiconductors, our results suggest that in the vicinity of that gap, the inter-atomic OHE reduces when increasing the band gap, qualitatively associated with the enhanced localization of the wave function. Conversely, in narrow-gap semiconductors the wave function becomes less localized close to the gap resulting in an enhanced inter-atomic OHE. We emphasize that the intra-atomic OHE is ubiquitous, except in specific cases where it is quenched by symmetry (gapped graphene \cite{Bhowal2021} or Fu's model \cite{Fu2011}). These results confirm that two-dimensional transition metals stand out as promising candidates for orbital current generation.

The obvious failure of the ACA in transition metals demonstrated above contrasts with previous theoretical studies \cite{Kontani2009,Jo2018,Salemi2022} and raises a number of questions, especially considering the importance of these materials for experiments. First, since light transition metals (V, Cr, Cu and their oxides) are considered as promising sources of OAM \cite{Ding2020,Lee2021b,Hayashi2022,Ding2022}, determining how much of this OAM is of intra- or inter-atomic origin is crucial for the realization of long-range orbital transport. The high sensitivity of OHE to band structure details and spin-orbit coupling calls for careful {\it ab initio} computations beyond the ACA.\par

Second, an important question that needs to be addressed is the distinct role of intra-atomic and inter-atomic contributions in orbital torque, pumping and magnetoresistance \cite{Go2020,Ding2020,Lee2021b,Ding2022}. As a matter of fact, these effects are all mediated by the spin-orbit coupling, which in most crystals reduces to the atomic Russel-Saunders coupling, $\sim \xi_{\rm so}{\bf S}\cdot{\bf L}$. Because the potential gradient responsible for the spin-orbit coupling is largest close to the nucleus, one can expect that the orbital torque, pumping and magnetoresistance mostly involve the intra-atomic OAM. Although this statement needs to be confirmed by precise theoretical calculations, it would mean that previous estimates of intra-atomic OHE \cite{Kontani2009,Jo2018,Salemi2022} could be used as guidelines for experimental design. This is an important question that, together with the distinct robustness against disorder of intra- and inter-atomic OHE, requires further experimental and theoretical investigations. 

\acknowledgments
This work was supported by the ANR ORION project, grant ANR-20-CE30-0022-01 of the French Agence Nationale de la Recherche. D. G.O. and A. M. acknowledge support from the Excellence Initiative of Aix-Marseille Universit\'e - A*Midex, a French "Investissements d'Avenir" program. The authors thank A. Saul, T. Rappoport, G. Vignale and D. Go for fruitful discussions.
	
\bibliography{refs-resub}

\end{document}